%% file: upswz_prd.tex
\documentclass[%
 reprint,
 amsmath,amssymb,
 aps,
 superscriptaddress,
 showpacs,
]{revtex4-1}

\newcommand{\MET}{\mbox{$E\kern-0.50em\raise0.10ex\hbox{/}_{T}$}}
\newcommand{\vecMET}{\mbox{$\vec{E}\kern-0.50em\raise0.10ex\hbox{/}_{T}$}}

\usepackage{graphicx}
\usepackage{dcolumn}
\usepackage{bm}

\begin{document}

\preprint{APS/123-QED}

\title{Search for production of an $\Upsilon \rm{(1S)}$ meson in association with a $W$ or $Z$ boson using the full 1.96 TeV 
$p\bar{p}$ collision data set at CDF}
 \input{author.tex}






\begin{abstract}
Production of the $\Upsilon {\rm (1S)}$ meson in association with a vector boson is a rare process 
in the standard model with a cross section predicted to be below the sensitivity of the Tevatron. 
Observation of this process could signify contributions not described by the standard model or reveal limitations with 
the current non-relativistic quantum-chromodynamic models used to calculate the cross section.
We perform a search for this process using the full Run II data set collected by the CDF II detector corresponding to an integrated
luminosity of $9.4\,{\rm fb^{-1}}$.
The search considers the $\Upsilon \rightarrow \mu\mu$ decay and the decay of the $W$ and $Z$ bosons into muons and electrons. 
In these purely leptonic decay channels, we observe one $\Upsilon W$ candidate with an expected background of $1.2 \pm 0.5$ 
events, and one $\Upsilon Z$ candidate with an expected background of $0.1 \pm 0.1$ events. Both observations are
consistent with the predicted background contributions. 
%
The resulting upper limits on the cross section for $\Upsilon + W/Z$ production are 
the most sensitive reported from a single experiment and place restrictions 
on potential contributions from non-standard-model physics.
%
%
\end{abstract}


\maketitle


\section{\label{sec:Intro}Introduction}

The standard model production of an upsilon ($\Upsilon$) meson in association with a $W$ boson or a $Z$ boson is a rare process 
whose rate was first calculated in Ref.~\cite{BLF}, where $\Upsilon W$ and  $\Upsilon Z$ production occur through the parton-level processes producing 
$W + b\bar{b}$ and $Z + b\bar{b}$ final states, in which the $b\bar{b}$ pair may form a bound state 
(either an $\Upsilon$ or an excited bottomonium state that decays to an $\Upsilon$).  More recently, rates for these processes have been calculated at 
next-to-leading-order in the strong-interaction coupling
for proton-antiproton ($p\bar{p}$) collisions at 1.96\,TeV center-of-mass energy and proton-proton collisions at 8\,TeV and 14\,TeV~\cite{YWZtheory} .


The cross sections calculated for $\Upsilon W$ and  $\Upsilon Z$ production in $p\bar{p}$ collisions at 1.96 TeV
are 43\,fb and 34\,fb, respectively. These values were calculated at leading-order using the {\sc Madonia}  quarkonium generator~\cite{madonia} as detailed  below
and are roughly a factor of ten smaller than the earlier calculations from Ref.~\cite{BLF}.
The calculations of these processes are very sensitive to the non-relativistic quantum-chromodynamic (NRQCD) models, 
especially the numerical values of the 
long-distance matrix elements (LDME), which determine the probability that a $b\bar{b}$ will form a bottomonium state. 
Measurements of  $\Upsilon + W/Z$ cross sections are important for validating these NRQCD models.

Supersymmetry (SUSY) is an extension of the standard model (SM) which has not been observed. Reference~\cite{BLF} describes
some SUSY models in which charged Higgs bosons can decay into $\Upsilon W$ final states with a large branching fraction (${\cal B}$). 
Similarly, in addition to the expected decays of a SM Higgs to an $\Upsilon Z$ pair, further light neutral scalars may decay into $\Upsilon Z$.
Therefore, if the observed rate of $\Upsilon W$  and/or $\Upsilon Z$  
production is significantly larger than the predicted SM rate, it may be an indication of physics not described by the SM.

In 2003, the CDF collaboration reported~\cite{Run1} a search for the associated production of an $\Upsilon$ meson and a $W$ or $Z$
boson. In that analysis, a sample corresponding to $83\,{\rm pb^{-1}}$ of 1.8 TeV $p\bar{p}$ collision data collected with the Run I CDF detector was used
to set upper limits on the production cross sections ($\sigma$) at the 95\% confidence level (C.L.) of 
$\sigma(p\bar{p} \rightarrow \Upsilon W) \times {\cal B}(\Upsilon \rightarrow \mu^+\mu^-) < 2.3\,{\rm pb}$ and
$\sigma(p\bar{p} \rightarrow \Upsilon Z) \times {\cal B}(\Upsilon \rightarrow \mu^+\mu^-) < 2.5\,{\rm pb}$. The ATLAS collaboration has also reported on the related channels of $J/\psi+W/Z$ production~\cite{DPSATLAS,Aad:2014kba}.


Here we present a search for  $\Upsilon + W/Z$  production, using a sample corresponding to $9.4\,{\rm fb^{-1}}$ of 1.96 TeV $p\bar{p}$ collision data collected with the CDF II detector.  
We use the dimuon decay channel to identify the $\Upsilon$ meson. 
We use only the electron and muon decay channels of the $W$ and $Z$ bosons, which give the best sensitivities for this search.

\section{The CDF Detector}

The CDF II detector is a nearly azimuthally and
forward-backward symmetric detector designed to study
$p\bar{p}$ collisions at the Tevatron. It is described in detail in
Ref.~\cite{CDF}. It consists of a magnetic spectrometer surrounded
by calorimeters and a muon-detection system. Particle
trajectories are expressed in a cylindrical coordinate
system, with the $z$ axis along the proton beam and the $x$ axis 
pointing outward from the center of the Tevatron.
The azimuthal angle ($\phi$) is defined with respect to the $x$ direction. 
The polar angle ($\theta$) is measured with respect to the $z$ direction, 
and the pseudorapidity ($\eta$) is defined as $\eta = -\ln(\tan{\frac{\theta}{2}})$.
The momentum of charged particles is measured by the
tracking system, consisting of silicon strip detectors surrounded
by an open-cell drift chamber, all immersed in a 1.4 T solenoidal magnetic field. 
The tracking system provides  charged-particle trajectory (track) information with good efficiency in the range  $|\eta| \lesssim 1.0$.
The tracking system is surrounded by pointing-geometry tower calorimeters,
that measure the energies of electrons, photons, and
jets of hadronic particles. The electromagnetic calorimeters
consist of scintillating tile and lead absorber,
while the hadronic calorimeters are composed
of scintillating tiles with steel absorber. The calorimeter system includes
central and plug subdetectors, with the
central region covering  $|\eta| < 1.1$
and the plug region covering the range $1.1 < |\eta| < 3.6$.
The muon system is composed of planar multi-wire drift
chambers. 
%
In the central region, four layers of chambers located just outside the calorimeter cover the region $|\eta| < 0.6$.  
An additional 60\,cm of iron shielding surrounds this system, and behind that is a second subdetector composed of another four layers of chambers.  
A third muon subdetector covers the region $0.6 < |\eta| < 1.0$, and a fourth subdetector extends coverage to $|\eta| < 1.5$.
Cherenkov luminosity counters measure the
rate of inelastic collisions, that is converted into the
instantaneous luminosity. 
A three-level online event-selection system (trigger) is used to reduce
the event rate from $2.5\,{\rm MHz}$ to  approximately $100\,{\rm Hz}$. The
first level consists of specialized hardware, while the second is a mixture
of hardware and fast software algorithms. The software-based
third-level trigger has access to a similar set of information to
that available in the offline reconstruction.

\section{Monte Carlo and Data Samples}

We use a number of quantities based on track and calorimeter information in the event selection. 
The transverse momentum of a charged particle is 
$p_T = p\sin\theta$, where $p$ is the particle's momentum. The analogous quantity measured
with the calorimeter is transverse energy, $E_T = E\sin\theta$.
The missing transverse energy, $\MET$ is defined as 
$\vecMET = -\sum_i E^i_T\hat{n}_i$, where $\hat{n}_i$ is a unit vector
perpendicular to the beam axis and pointing to the center of the $i$th
calorimeter tower. The $\vecMET$ is adjusted for high-energy muons, that deposit only a small fraction of their energies in the calorimeter,
and offline corrections applied to the measured energies of reconstructed jets~\cite{jet_corrections} which result from the hadronization of quarks and gluons.
We define $\MET = |\vecMET |$. The invariant mass
of two leptons is $M_{\ell\ell} = \sqrt{(E_{\ell 1} + E_{\ell 2})^2/c^4 - |\vec{p}_{\ell1} + \vec{p}_{\ell2}|^2/c^2}$, and the transverse mass of a lepton and neutrino (estimated with $\MET$) is 
$M_T = \sqrt{2E_T^{\ell}\MET (1 - \cos\xi)/c^3}$ where $\xi$ is the angle between the lepton track and $\MET$ vector in the
transverse plane. For muons, $p_{\ell}$ and $p_T^{\ell}$ are used rather than their measured energies $E_{\ell}$ and $E_T^{\ell}$ in the definitions of $M_{\ell\ell}$ and $M_T$.

The analysis uses events 
selected with triggers requiring a high-$E_T$ central electron candidate ($E_T > 18\,{\rm GeV}$, $|\eta | < 1.0$)
or a high-$p_T$ central muon candidate ($p_T > 18\,{\rm GeV}/c$, $|\eta | < 1.0$).
The integrated luminosity of these samples is $9.4\,{\rm fb^{-1}}$. 
All the search channels include the $\Upsilon \rightarrow \mu\mu$ signal, so we only use data acquired when the muon detectors were operational, 
resulting in the same integrated luminosity for the electron and muon samples. 

\begin{figure}[htbp]
  \begin{center}
    \includegraphics[width=\columnwidth]{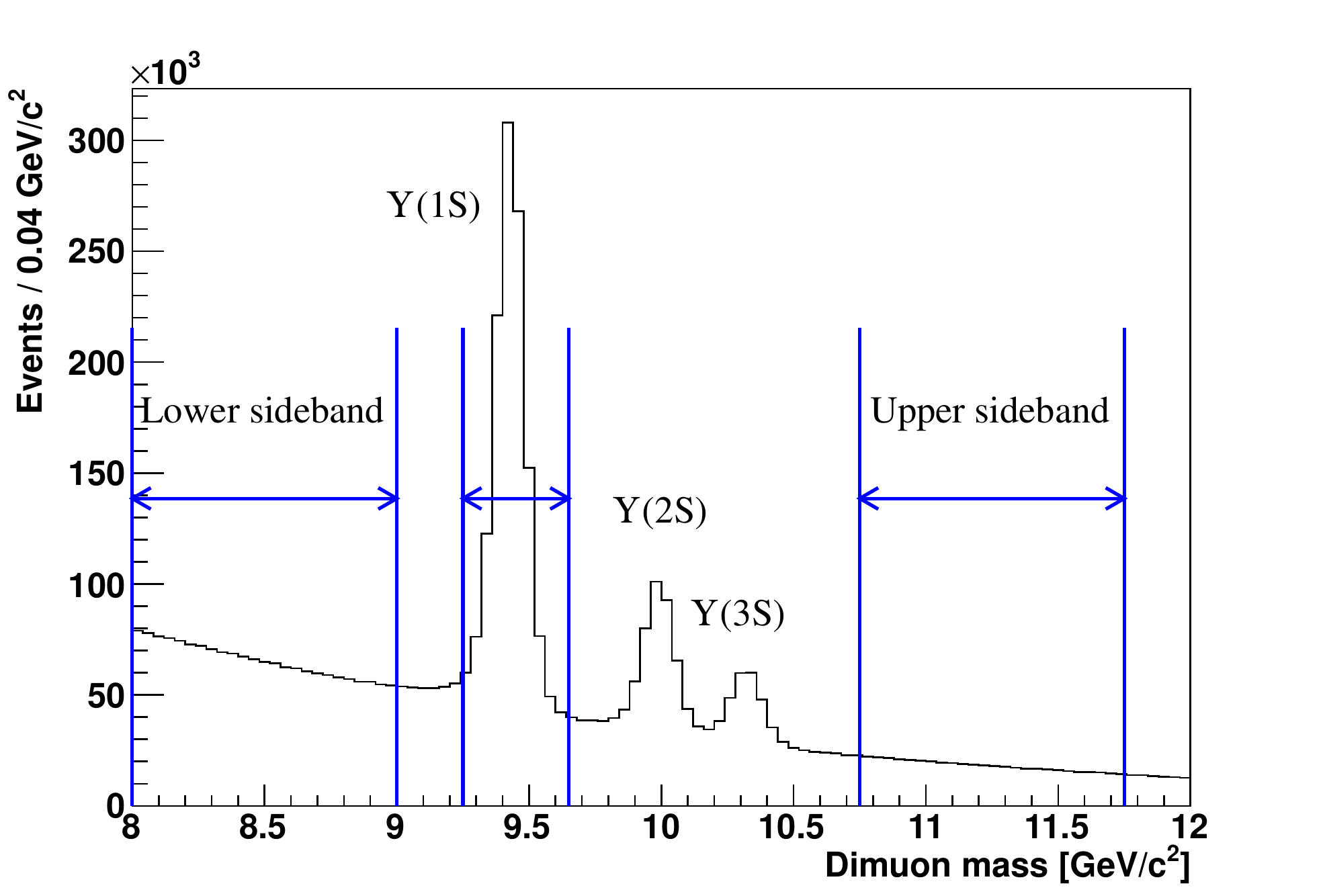}
    \caption{Dimuon invariant-mass spectrum in CDF II data from events contained within the low-$p_T$ dimuon-triggered sample. Shown
      are the defined $\Upsilon$ signal region and the sideband regions used for background determination. \label{fig:mm_spectrum}} 
\end{center}
\end{figure}

We also use a low-$p_T$ dimuon-triggered $\Upsilon$ sample to estimate one of the backgrounds as detailed in Section~\ref{backgrounds}.
The dimuon invariant-mass distribution from this low-$p_T$ sample, whose integrated luminosity is $7.3\,{\rm fb^{-1}}$, is shown
in Fig.~\ref{fig:mm_spectrum} for the mass range in the region of the $\Upsilon$ resonances.

%
We produce simulated event samples of the signal processes, $\Upsilon W$ and $\Upsilon Z$, by first generating events with {\sc Madgraph}~\cite{madgraph} and 
its quarkonium extension {\sc Madonia}~\cite{madonia}.
We include all $\Upsilon W$ and $\Upsilon Z$ processes from Ref.~\cite{BLF} and the LDME values relevant for the Tevatron from 
Ref.~\cite{LDME}.  
An explanation of how LDME values are determined from fits to quarkonia data is given in Ref.~\cite{LDME2}, although the 
values obtained in this reference are specific for the LHC.
{\sc Pythia}~\cite{pythia} is used to simulate the $\Upsilon$, $W$, and $Z$ decays and parton showering. 
Generated $\Upsilon$ mesons are forced to decay to two muons. 
We use a {\sc Geant3}-based~\cite{geant} detector simulation to model the response of the CDF II detector~\cite{cdfsim}.

\section{Event Selection}
\label{event_selection}

Events are selected with $\Upsilon$ mesons decaying to muon pairs and decays of vector bosons
resulting in at least one electron or muon.
In this analysis we have two categories of lepton candidates: low-$p_T$ muon candidates with $1.5 < p_T < 15\,{\rm GeV}/c$ from the $\Upsilon$ decay and 
high-$E_T$ (or $p_T$) 
electron (or muon) candidates from the $W$ or $Z$ decay. 


High-$E_T$ electron candidates are identified by matching a 
track to energy deposited within the calorimeter.
Muon candidates are formed from charged particle tracks matched to minimum ionizing energy deposition 
in the calorimeter, which may or may not be matched 
to track segments in the muon chambers situated behind the calorimeters.
Lepton reconstruction algorithms are described in detail elsewhere~\cite{Abulencia:2005ix}.

Electron candidates are distinguished by whether they are found in the central or forward calorimeters ($|\eta|>1.1$)
where only silicon tracking information is available. The electron selection relies on track quality, track-calorimeter 
matching, calorimeter energy, calorimeter profile shape, and isolation information.
Most muon candidates rely on direct detection in the muon chambers, which are
distinguished by their acceptance in pseudorapidity: central muon detectors ($|\eta|<0.6$), central muon extension detectors ($0.6<|\eta|<1.0$),  and the intermediate muon detector ($1.0<|\eta|<1.5$). Remaining muon candidates rely on track matches to energy deposits consistent with a minimum
ionizing charged particle in the central and forward electromagnetic calorimeters respectively, and which fail to have an 
associated track segment in the muon sub-detectors.
All high-$E_T$ (or $p_T$) leptons are required to be isolated by imposing the condition that 
the sum of the transverse energy of the calorimeter towers in a cone of 
$\Delta R \equiv \sqrt{(\Delta\phi)^2+(\Delta\eta)^2}=0.4$
around the lepton is less than $10$\% of the electron $E_T$ (muon $p_T$). 

The analysis uses the high-$E_T$ electron triggered, and high-$p_T$ muon triggered, data sets where events are additionally
required to contain $\Upsilon (1S)$ candidates using the $\Upsilon$ decay to two low-$p_T$ muons 
($1.5 < p_T < 15\,{\rm GeV}/c$).
 We define the $\Upsilon$(1S) region as the invariant-mass range $9.25 < M_{\mu\mu} < 9.65\,{\rm GeV}/c^{2}$. We do not use
$\Upsilon$(2S) or $\Upsilon$(3S) decays.
We define two sideband regions,  $8.00 < M_{\mu\mu} < 9.00\,{\rm GeV}/c^{2}$ 
and $10.75 < M_{\mu\mu} < 11.75\,{\rm GeV}/c^{2}$, for obtaining background estimates.
Events are required to have at least two low-$p_T$ muon candidates
whose invariant mass lies within the $\Upsilon$(1S) region. To increase the efficiency for reconstructing $\Upsilon$ candidates, we
use looser quality requirements on these low-$p_T$  muon candidates than for the high-$p_T$ muon candidates
used in the vector-boson reconstruction. 
In particular, there are no isolation requirements on the $\Upsilon$ muon candidates, 
and geometrical matching requirements between charged particles in the tracker and track segments in the muon detectors are
less stringent. Most low-$p_T$ muon candidates surviving event selection are found to be within acceptance of the muon chambers ($|\eta|<1.5$).
In the small fraction of events (less than 2\%) that have more than two low-$p_T$ muons identified, we randomly choose one pair of those muons.

\begin{figure}[htbp]
\begin{center}
\includegraphics[width=\columnwidth]{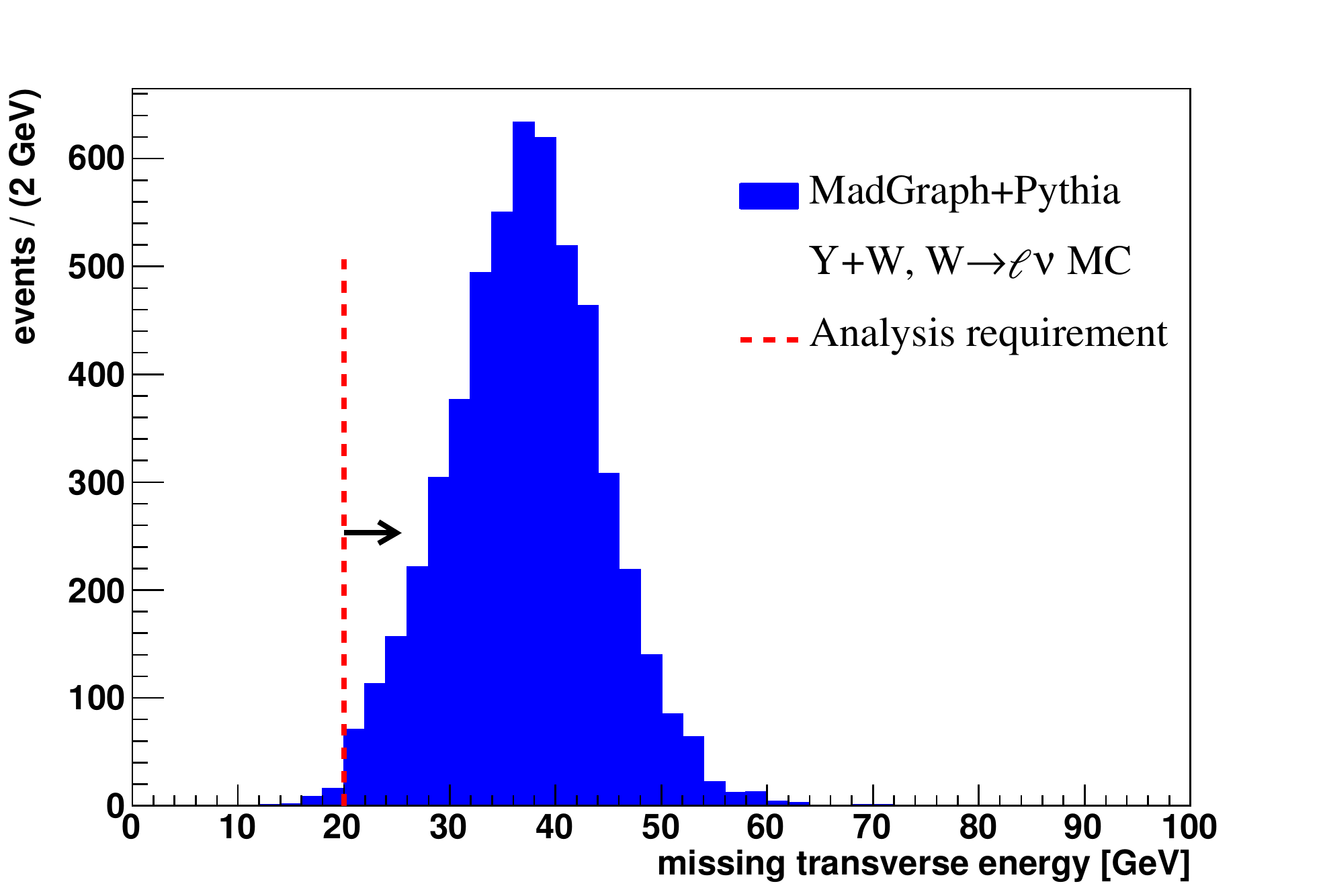}
\caption{Missing-transverse-energy distributions predicted for signal  $\Upsilon + (W \rightarrow \ell\nu)$ events.  
The distributions are shown  for events that satisfy all other event requirements. The scale of the vertical axis is arbitrary.  }
\label{fig:met}
\end{center}
\end{figure}
\begin{figure}[htbp]
\begin{center}
\includegraphics[width=\columnwidth]{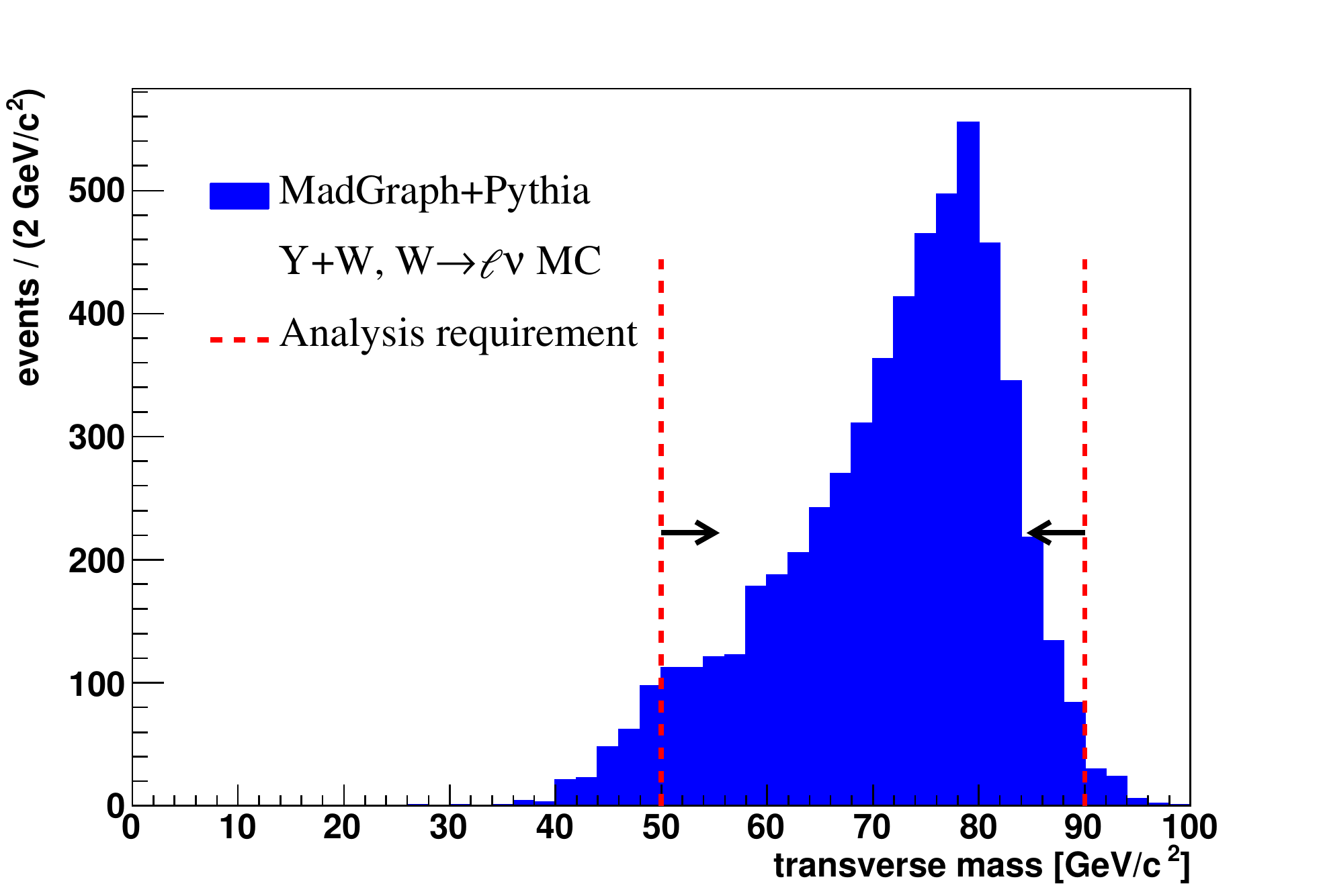}
\caption{Transverse-mass distributions predicted for signal  $\Upsilon + (W \rightarrow \ell\nu)$ events.  
The distributions are shown  for events that satisfy all other event requirements. The scale of the vertical axis is arbitrary.  }
\label{fig:mt}
\end{center}
\end{figure}

We then look for additional high-energy
electron (or muon) candidates consistent with the decay of a vector boson. 
Events with exactly one high-energy lepton candidate, $\ell$, which will henceforth refer to an electron or muon,
with $E_T$ ($p_T$) greater than 20\,GeV\,(GeV/$c$),
in addition to the $\Upsilon \rightarrow \mu^+ \mu^-$ candidate, and significant missing transverse energy 
($\MET > 20\,{\rm GeV}$) are selected as
$\Upsilon + (W \rightarrow \ell \nu)$ candidates. Such candidates are further required to have a transverse mass in the range 
$50 < M_T < 90\,{\rm GeV}/c^{2}$, as expected from a $W$ boson decay.
Figure~\ref{fig:met} and Figure~\ref{fig:mt} show the distributions of these quantities as predicted from the simulated 
$\Upsilon + W$ event samples.

\begin{figure}[htbp]
\begin{center}
\includegraphics[width=\columnwidth]{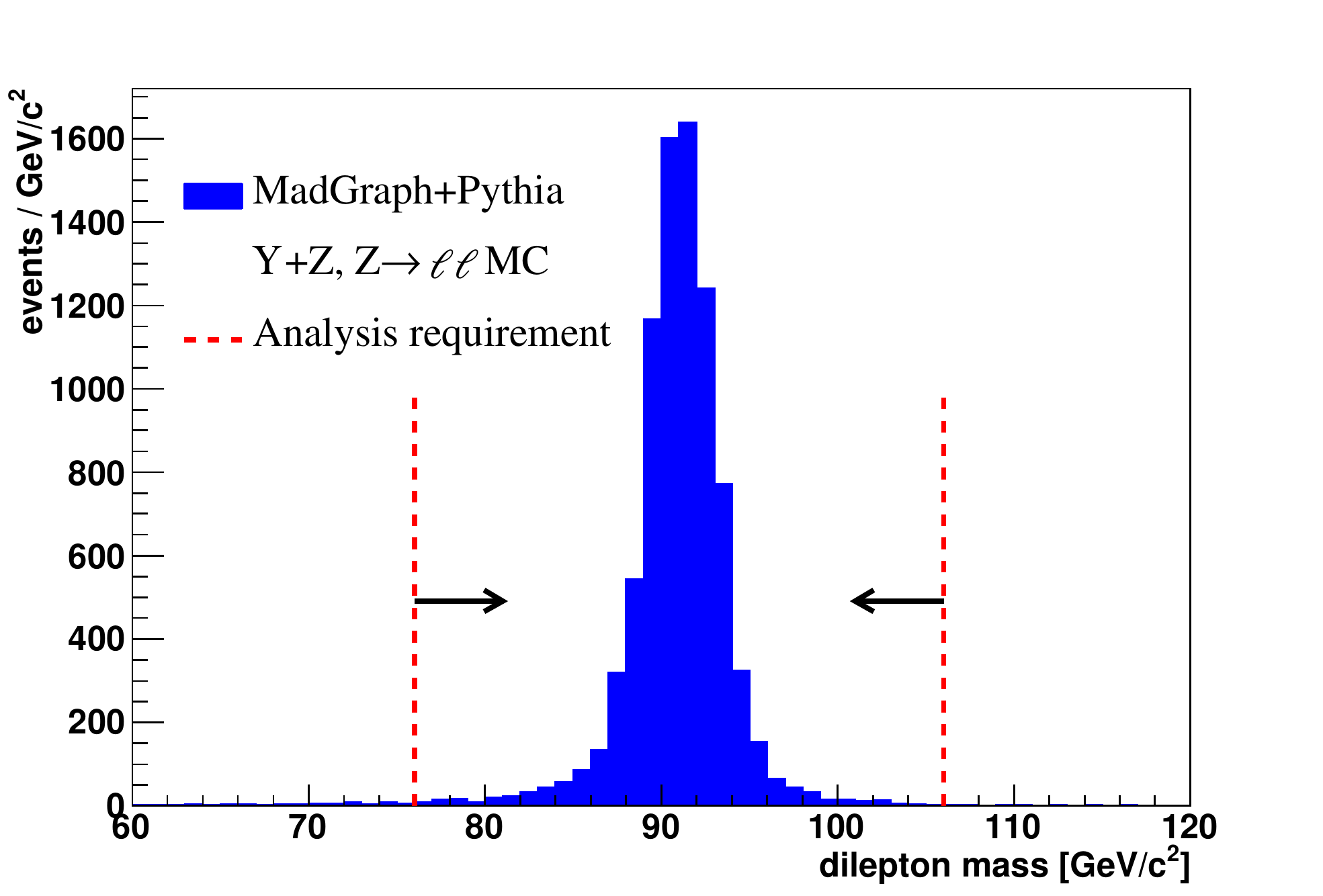}
\caption{Dilepton invariant-mass distribution predicted for signal  $\Upsilon + (Z \rightarrow \ell\ell)$ events. 
The distribution is shown  for events that satisfy all other event requirements. The scale of the vertical axis is arbitrary.} 
\label{fig:mll}
\end{center}
\end{figure}

Events with two oppositely charged high-energy
lepton candidates of same flavor are selected as $\Upsilon + (Z \rightarrow \ell^+ \ell^-)$ candidates.
The $\Upsilon + (Z \rightarrow \ell\ell)$ candidates are selected by 
requiring one additional high-$E_T$($p_T$) electron (muon) candidate with $E_T (p_T) > 20\,{\rm GeV}\,{\rm (GeV}/c)$ and a second candidate with the same 
flavor but opposite charge and $E_T (p_T) > 15\,{\rm GeV}\,{\rm (GeV/}c)$. 
Both additional lepton candidates are required to be isolated and have an invariant mass in the range $76 < M_{\ell\ell} <106\,{\rm GeV}/c^{2}$. 
The invariant-mass distribution predicted from the simulated $\Upsilon + (Z \rightarrow \ell\ell)$ event samples is shown in Fig.~\ref{fig:mll}.

The total signal efficiencies, after all selection criteria are applied, are determined from simulated event samples to be 1.8\% for $\Upsilon + (W \rightarrow e\nu)$, 
1.3\% for  $\Upsilon + (W \rightarrow \mu\nu)$, 1.8\% for  $\Upsilon + (Z \rightarrow ee)$, and 1.4\% for
$\Upsilon + (Z \rightarrow \mu\mu)$ events. These efficiencies do not include the branching fractions for $\Upsilon \rightarrow \mu\mu$
and the electronic and muonic decays of the vector bosons. The low acceptances are primarily driven by the geometric acceptance of the drift chamber 
for the two low-$p_T$ muons from the $\Upsilon$ decay. We expect a small contribution to the $W\rightarrow \ell\nu$ acceptance
from $W \rightarrow \tau\nu$ events where the tau lepton decays to an electron or muon. The contribution is determined to 
be less than 2\% of the acceptance, and is therefore neglected. The contribution from $Z \rightarrow \tau\tau$ events to the 
$Z \rightarrow \ell\ell$ channels is found to be negligible.

Summaries of the selection criteria and their associated efficiencies are given in Tables~\ref{efficiency1} and~\ref{efficiency2}.

\begin{table*}[ht]
  \caption{Efficiencies for the $(\Upsilon \rightarrow \mu\mu) + (W \rightarrow \ell\nu)$ selection criteria.
The individual efficiencies for each requirement, in the given order, are listed with the total at the bottom.
  The uncertainty on the total efficiency is discussed in the text.}
  \begin{center}
    \begin{tabular}{ l  c  c  r }
    \hline
    \hline
      & \ \ $\Upsilon + (W\rightarrow e\nu)$\ \  & \ \ $\Upsilon + (W\rightarrow \mu\nu)$\ \  \\
    \hline
    $\Upsilon (1S) \rightarrow \mu\mu$ candidate & 6.8$\%$ & 6.8$\%$ \\
    One additional high-$E_T$($p_T$) isolated $e$ or $\mu$ candidate & 55$\%$ & 46$\%$ \\
    High-$E_T$($p_T$) lepton candidate is triggerable & 55$\%$ & 52$\%$ \\
    $\MET > 20\,{\rm GeV}$ & 96$\%$ & 94$\%$ \\
    $50 < M_T < 90\,{\rm GeV}/c^{2}$ & 94$\%$ & 95$\%$ \\
    Trigger efficiency & 97$\%$ & 92$\%$ \\
    \hline
    Total & $(1.8 \pm 0.4)\%$ & $(1.3\pm 0.3)\%$ \\
    \hline
    \hline
    \end{tabular}
  \end{center}
  \label{efficiency1}
\end{table*}

\begin{table*}[ht]
  \caption{Efficiencies for the $(\Upsilon \rightarrow \mu\mu) + (Z \rightarrow \ell\ell)$ selection criteria. 
 The individual efficiencies for each requirement, in the given order, are listed with the total at the bottom.
  The uncertainty on the total efficiency is discussed in the text.
  OS means opposite-sign.}
  \begin{center}
    \begin{tabular}{ l  c  c  r }
    \hline
    \hline
      & \ \ $\Upsilon + (Z\rightarrow ee)$ \ \ & \ \ $\Upsilon + (Z\rightarrow \mu\mu)$\ \  \\
    \hline
    $\Upsilon (1S) \rightarrow \mu\mu$  candidate & 6.7$\%$ & 7.0$\%$ \\
    Two additional OS high-$E_T$($p_T$) isolated $e$ or $\mu$ candidates & 32$\%$ & 25$\%$ \\
    One of the two high-$E_T$($p_T$) lepton candidates is triggerable & 86$\%$ & 80$\%$ \\
    $76 <  M_{\ell\ell} < 106\, {\rm GeV}/c^{2}$ & 99$\%$ & 99$\%$ \\
    Trigger efficiency & 98$\%$ & 95$\%$ \\
    \hline
    Total & $(1.8\pm 0.4)\%$ & $(1.4\pm 0.3)\%$ \\
    \hline
    \hline
    \end{tabular}
  \end{center}
  \label{efficiency2}
\end{table*}

\section{Backgrounds}
\label{backgrounds}

There are two main background contributions to the samples of
$\Upsilon W$ and $\Upsilon Z$ signal candidates after the final selection:
events containing a correctly identified $W/Z$ candidate 
and a misidentified $\Upsilon$ candidate (real $W/Z$ + fake $\Upsilon$) 
and those with a correctly identified $\Upsilon$ candidate and a 
misidentified $W/Z$ candidate (real $\Upsilon$ + fake $W/Z$).  Generic 
dimuon backgrounds, originating predominantly from $b\bar{b}$
production, contribute events in the $\Upsilon {\rm (1S)}$ mass range and are the primary source of fake 
$\Upsilon$ candidates.  Misidentification of jets as leptons 
can mimic the decay signatures of $W$ and $Z$ bosons.  In the case of $Z$ 
candidates, where two leptons are required, this background is negligible.

The real $W/Z$ + fake $\Upsilon$ background contributions are
estimated by counting the number of $W$ or $Z$ candidate events  in the high-$p_T$ lepton data samples
that additionally contain a dimuon candidate in the sideband region of the dimuon spectrum 
(defined in Fig.~\ref{fig:mm_spectrum}).
An exponential fit to these sideband regions is used to determine a ratio of the areas of the signal to
sideband regions, which is then applied to these numbers for an estimate of this  background contribution. 


The probabilities for reconstructed jets to be misidentified as leptons are measured in jet-enriched data samples as 
functions of the jet $E_T$ and lepton type, and are corrected for the contributions of leptons from $W$ and $Z$ boson decays, as more fully described in Ref.~\cite{HWW}.
To estimate real $\Upsilon$ + fake $W/Z$ background
contributions, we select from the low-$p_T$ dimuon data sample 
events containing a high-$E_T$ 
jet instead of a high-$E_T$ 
($p_T$) isolated lepton candidate that otherwise satisfy the 
full selection criteria.  Background estimates are obtained 
using the measured probabilities associated with each of the 
jets within these events as weighting factors 
on the potential contribution of each.  The low-$p_T$ dimuon 
sample is relied upon to extract these background estimates 
because a strong correlation between high-$p_T$ lepton trigger 
selection requirements and jet-to-lepton misidentification 
rates renders the high-$p_T$ lepton data set unsuitable for the 
chosen methodology.  To interpolate between the two samples,
additional small corrections are applied to account 
for differences in the integrated luminosities of the two 
samples and $\Upsilon$ selection inefficiencies in the low-$p_T$
dimuon sample originating from trigger requirements.             

The predicted background contributions to each of the signal 
samples are summarized in Table~\ref{summary_all}.  In evaluating the real $Z$ 
+ fake $\Upsilon$ background contribution, no events containing 
$\Upsilon$ candidates in the sideband mass regions 
are observed.  Background contributions to the corresponding 
signal samples are therefore estimated by extrapolating from 
the estimated real $W$ + fake $\Upsilon$ background contributions,
using the ratio of $Z$-to-$W$ cross sections. This makes the assumption
that the probability for misidentifying a $\Upsilon \rm{(1S)}$ is independent of the type of vector boson.  
In calculating cross-section limits, we also 
account for small background contributions from $\Upsilon Z$ 
production to the $\Upsilon W$ samples, originating 
from events in which one of the two leptons produced in the 
$Z$ boson decay is not reconstructed.          

\section{Systematic Uncertainties}

For determining cross-section limits we incorporate systematic uncertainties on the signal expectation 
and the background predictions.
Systematic uncertainties on the signal expectation include those associated 
with the  integrated luminosity measurement, low-$p_T$ muon identification, 
high-$E_T$($p_T$) lepton identification, high-$E_T$($p_T$) lepton trigger efficiency, theoretical modeling of the signal, and efficiencies 
of the event selection criteria. 
 The upsilon-muon identification uncertainty is derived from studies that use data and simulated samples of $J/\psi \to \mu\mu$ 
 as described in Ref.~\cite{Ypol}.  Lepton identification and trigger efficiencies are measured using samples of leptonic $Z$ 
 decays~\cite{HWW}. Requirements of $E_T > 20$\,GeV ($p_T > 20$\,GeV/$c$) for electrons (muons) matched to lepton trigger objects ensure a uniform trigger efficiency over the lepton momentum spectra.

We use the CTEQ6L parton distribution functions (PDFs)~\cite{cteq} for generating the {\sc Madgraph} samples.  To estimate the acceptance uncertainty associated with the choice
of PDFs, we generate additional samples using MRST PDFs~\cite{mrst} and take the difference in the estimated signal acceptance as the uncertainty. 

We vary the bottomonium LDMEs from Ref.~\cite{LDME} by one standard deviation to estimate their effect on the 
signal acceptance.
This procedure results in an additional 6\% systematic uncertainty on the signal acceptance. 
These uncertainties correspond only to those associated with the procedure 
for computing LDMEs described within the cited reference.  Allowing for a  
wider range of assumptions within the LDME calculations gives rise to 
additional uncertainties, which are not accounted for in this analysis.
However, if an uncertainty of 20\% were to be placed on the LDMEs, the cross-section limits we obtain would only increase by about 10\%. 

With respect to uncertainties associated with event selection criteria, we vary the $\MET$ by $\pm 10\%$ (an estimate of the $\MET$ resolution) in the simulated signal 
samples to quantify the effect of $\MET$ resolution.


It is possible for the $\Upsilon$ meson and the $W$ or $Z$ boson to originate from different parton-parton interactions in the 
same $p\bar{p}$ collision.
This double-parton-scattering process is difficult to identify, but estimates have been made for several related final states
using LHC and Tevatron data (see for example Ref.~\cite{DPSATLAS} where $J/\psi$ production in association with a $W$
boson was studied by the ATLAS collaboration).
These estimates, together with a calculation using the $\Upsilon$ and vector boson cross sections at the Tevatron collision energy
lead to an estimated effect of approximately 15\%. Based on lack of knowledge on double-parton scattering, we assign this 
effect as a systematic uncertainty on the signal acceptance.
In Table~\ref{signal_systematics} we summarize all investigated systematic uncertainties associated with the signal expectation.


Uncertainties on predicted background contributions are also incorporated into the cross-section limits.
For  the real $W/Z$ + fake $\Upsilon$ background, we use the statistical uncertainty  originating from the
small sample size in the sideband regions used for making this estimate.  We assign a 50\% uncertainty to the real 
$\Upsilon$ + fake $W/Z$ background based on the application of uncertainties associated with the measured jet-to-lepton misidentification rates.

 \begin{table}[ht]
 \caption{Systematic uncertainties associated with the signal expectation.}
  \begin{center}
    \begin{tabular}{ l  c }
    \hline
    \hline
    Luminosity & 6\% \\
    $\Upsilon$ muon identification & 4\% \\
    High-$E_T$($p_T$) lepton identification & 1\% \\
    High-$E_T$($p_T$) lepton trigger efficiency \ \ \ & 1\% \\
    PDFs & 12\% \\
    LDMEs & 6\% \\
    Double parton scattering & 15\% \\
    Event selection efficiency & 3\% \\
    \hline
    Total &  22\% \\
    \hline
    \hline
    \end{tabular}
  \end{center}
  \label{signal_systematics}
\end{table}

\section{Results}

Table~\ref{summary_all} summarizes the predicted signal and background contributions, and number of observed events for each
of the search samples using data from $9.4\,{\rm fb^{-1}}$ of integrated luminosity at CDF. 
We observe one $\Upsilon + (W\rightarrow\ell\nu)$ candidate with a total expected background of
$1.2 \pm 0.5$ events.
In the observed $\Upsilon + (W\rightarrow\ell\nu)$ candidate the electron has $p_T = 27.4$\,GeV, and the two muons 
with an invariant mass in the $\Upsilon$(1S) region have $p_T$s of $3.8\,{\rm GeV}/c$ and $7.1\,{\rm GeV}/c$. The $\MET$ in
this event is $30.8\,{\rm GeV}$, which, with the electron gives a transverse mass of $58.1\,{\rm GeV}/c^2$.

\begin{figure}[htbp]
  \begin{center}
    \includegraphics[width=\columnwidth]{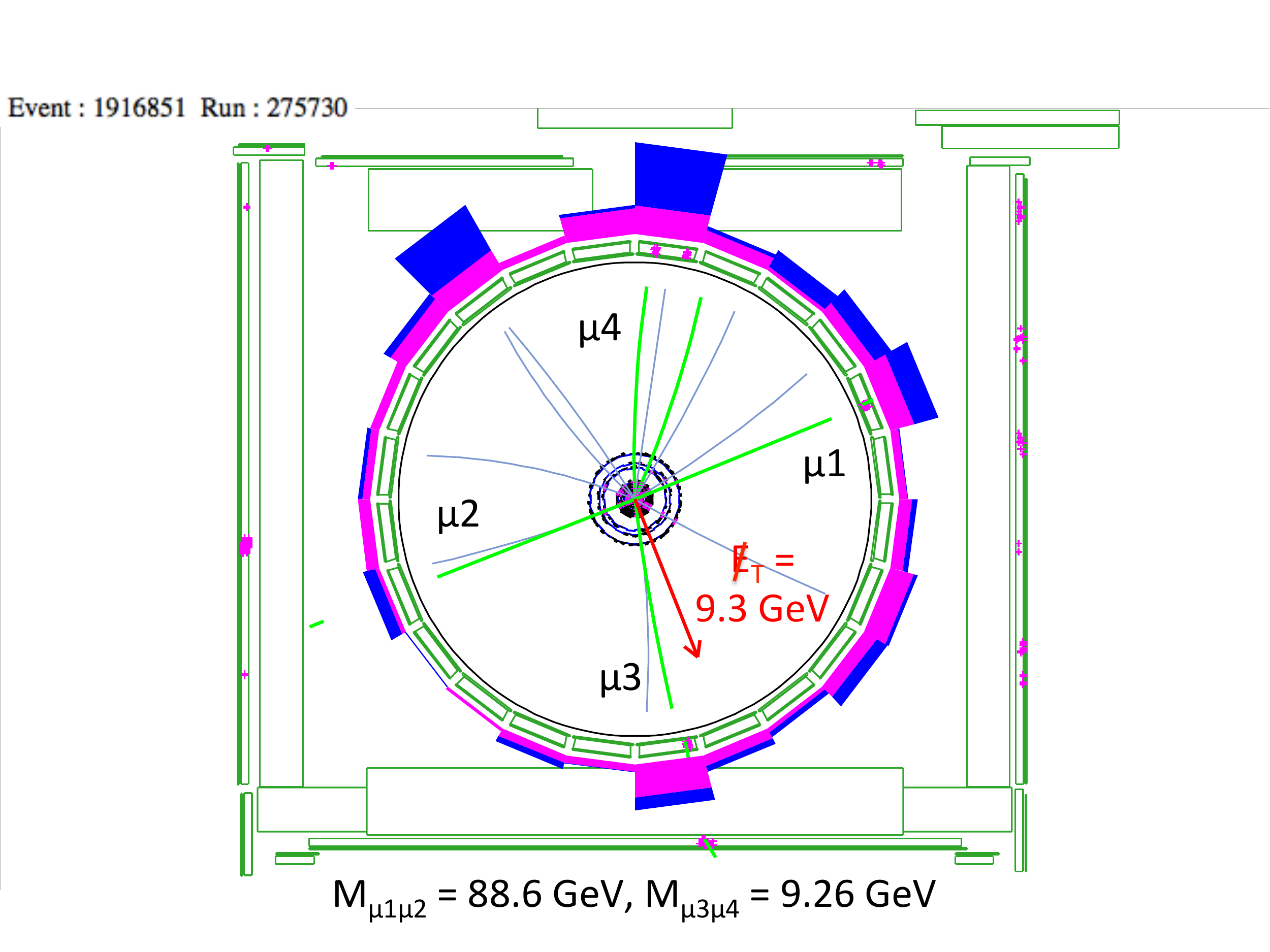}
    \caption{Event display of the observed $\Upsilon Z$ candidate, showing the muon candidates identified from the $\Upsilon$ 
     and $Z$ decays. The view is in the transverse ($r - \phi$) plane of the detector, where the inner core is the silicon
     vertex tracker, and the larger circle is the outer radius of the drift chamber where the tracks of charged particle with 
     $p_T > 1.5\,{\rm GeV}/c$ are shown. The height of the surrounding pink and blue ``towers" is roughly proportional to the 
     energy deposits in the electromagnetic and hadronic compartments of the calorimeter, from which the $\MET$ magnitude 
     and direction (red arrow) is computed. Measurement-hits in the muon chambers are shown in the outermost box-shaped
     structure.
     \label{fig:evd} }
  \end{center}
\end{figure}

We also observe one $\Upsilon + (Z\rightarrow\ell\ell)$ candidate with a total expected background of $0.1 \pm 0.1$
events. An event display of the  $\Upsilon + (Z\rightarrow\ell\ell)$ candidate is shown in Fig.~\ref{fig:evd}. This is the first observed
$\Upsilon + (Z\rightarrow\ell\ell)$ candidate event at the Tevatron. The two high-$p_T$ muon candidates have an invariant mass of $88.6\, {\rm GeV/}c^{2}$, and the
two low-$p_T$ muon candidates have an invariant mass of $9.26\, {\rm GeV/}c^{2}$. All muon candidates are detected in the central region of the detector. 
The invariant mass of all four muon candidates is $98.4\, {\rm GeV/}c^{2}$. Further properties of the muons in this event are 
given in Table~\ref{golden_event_info}.

\begin{table*}[ht]
 \caption{Summary of signal expectation (N$_{sig}$), background estimations (N$_{bg}$), and observed events (N$_{obs}$).
  }
  \begin{center}
    \begin{tabular}{ l  c  c  c  c  c  c }
    \hline
    \hline
      & $\Upsilon + W\rightarrow e\nu$ & $\Upsilon + W\rightarrow \mu\nu$ & $\Upsilon + W\rightarrow\ell\nu$ & 
       $\Upsilon + Z\rightarrow ee$ & $\Upsilon + Z\rightarrow \mu\mu$ & $\Upsilon + Z\rightarrow \ell\ell$ \\
    \hline
    N$_{sig}$ & 0.019$\pm$0.004 & 0.014$\pm$0.003 & 0.034$\pm$0.007 & 0.0048$\pm$0.0011 & 0.0037$\pm$0.0008 & 0.0084$\pm$0.0018 \\
    \hline
    N$_{bg}$ (fake $\Upsilon$) & 0.7$\pm$0.4 & 0.4$\pm$0.3 & 1.1$\pm$0.5 & 0.07$\pm$0.07 & 0.04$\pm$0.04 & 0.1$\pm$0.1 \\

    N$_{bg}$ (fake $W/Z$) & 0.06$\pm$0.04 & 0 & 0.06$\pm$0.04 & 0 & 0 & 0\\

    N$_{bg}$ ($\Upsilon + Z$) & 0.0006$\pm$0.0001 & 0.0033$\pm$0.0007 & 0.0039$\pm$0.0009 & & & \\
    \hline
    N$_{bg}$ (total) & 0.8$\pm$0.4 & 0.4$\pm$0.3 & 1.2$\pm$0.5 & 0.07$\pm$0.07 & 0.04$\pm$0.04 & 0.1$\pm$0.1  \\
    \hline
    N$_{obs}$ & 0 & 1 & 1 & 0 & 1 & 1  \\
    \hline
    \hline
    \end{tabular}
  \end{center}
  \label{summary_all}
\end{table*}

Having observed no clear evidence for a $\Upsilon + W/Z$ signal, we set 90\% C.L. and 95\% C.L.  upper limits on the 
 $\Upsilon W$ and  $\Upsilon Z$ production cross sections. 
 We use the branching fractions of $\Upsilon \rightarrow \mu\mu$ (0.0248), $W \rightarrow \ell\nu$ (0.107), and $Z \rightarrow \ell\ell$ (0.0336) from
 Ref.~\cite{PDG}.
 A Bayesian technique is employed, described in Ref.~\cite{limits}, where the posterior probability density was constructed from the joint Poisson probability of observing
 the data in each vector boson decay channel, integrating over the uncertainties of the normalization parameters using Gaussian prior-probability densities. 
 A non-negative constant prior in the signal rate was assumed.
The expected and observed limits are shown in Table~\ref{limits} and compared to the observed limits from the CDF 
Run I analysis~\cite{Run1}. 




\begin{table}[ht]
\caption{Cross-section upper limits for $\Upsilon W$ and $\Upsilon Z$ production. This analysis utilizes $9.4\, {\rm fb^{-1}}$ of CDF Run II data. The CDF Run I analysis utilized $83\,{\rm pb^{-1}}$ of CDF Run I data. }
  \begin{center}
    \begin{tabular}{ l  c  c }
    \hline
    \hline
      & \ \ \ $\Upsilon W$\ \ \  & \ \ \  $\Upsilon Z$\ \ \  \\
\hline
    90\% C.L. expected limit (pb)\ \ \ \  & 4.4 & 9.9 \\
    90\% C.L. observed limit (pb) & 4.4 & 16 \\
    \hline
    95\% C.L. expected limit (pb) & 5.6 & 13 \\
    95\% C.L. observed limit (pb) & 5.6 & 21 \\
    \hline
    Run I 95\% C.L. observed limit (pb) \ \ & 93 & 101 \\
    \hline
    \hline
    \end{tabular}
  \end{center}
  \label{limits}
\end{table}

\begin{table}[ht]
\caption{Kinematic properties of the muons in the observed $\Upsilon +Z$ candidate displayed in Fig.~\ref{fig:evd}.
    Isolation is defined as the sum of calorimeter energy in a cone of $\Delta R = 0.4$ around the muon candidate as a fraction of the 
    muon momentum. The longitudinal position $z_0$ (along the beam line) of each muon candidate suggest all muons come from
     the same primary $p\bar{p}$ interaction vertex. }
  \begin{center}
    \begin{tabular}{ l | c  c  c  c }
    \hline\hline
      &\ \ \ muon 1\ \ \  & \ \ \ muon 2\ \ \  & \ \ \ muon 3\ \ \  & \ \ \ muon 4\ \ \  \\ 
    \hline
    $p_x$ (GeV/$c$)\ \ \  & -34.6 & 34.8 & 0.823 & -0.106 \\
    $p_y$ (GeV/$c$) & -14.0 & 13.8 & -6.25 & 3.29 \\
    $p_z$ (GeV/$c$) & -39.2 & 10.6 & -2.20 & -2.56 \\    
    $E$ (GeV) & 54.2 & 39.0 & 6.68 & 4.17 \\
    $p_T$ (GeV/$c$) & 37.4 & 37.5 & 6.3 & 3.3 \\
    $\eta$ & -0.92 & 0.28 & -0.34 & -0.72 \\
    $\phi$ (rads) & -2.76 & 0.38 & -1.44 & 1.60 \\
    Isolation & 0.03 & 0.00 & 0.64 & 0.35 \\
    $z_0$ (cm) & 41.2 & 41.1 & 41.0 & 41.3 \\
   \hline \hline
    \end{tabular}
  \end{center}
    \label{golden_event_info}
 \end{table}

\section{Conclusions}

We search for $\Upsilon + W/Z$
production using the leptonic decay channels of the
vector bosons and dimuon decay channel of the $\Upsilon$. The search
utilizes the full CDF Run II data set. Having observed no significant 
excess of events with respect to standard model predictions, we set 95\% C.L. upper limits on the $\Upsilon + W/Z$
cross sections. The limits are $\sigma(p\bar{p} \rightarrow \Upsilon W)  <  5.6\,{\rm pb}$ and 
 $\sigma(p\bar{p} \rightarrow \Upsilon Z)  <  21\,{\rm pb}$
which are the most stringent bounds on these processes to date.
Under the assumption that potential non-SM physics contributions to the $\Upsilon + W/Z$
final state do not significantly impact the kinematic properties  of  events, these limits
can be interpreted as  cross section (times branching ratio to $\Upsilon + W/Z$)
limits on non-SM physics processes contributing to this final state. Potential non-standard-model heavy particles decaying to 
$\Upsilon + W/Z$ final states
are likely to result in leptons that are more central than those from standard-model $\Upsilon + W/Z$
production and therefore provide higher signal acceptance.  Hence, the limits presented here can be considered as conservative limits on such processes.

\begin{acknowledgments}

We would like to acknowledge K.~W.~Lai for suggesting the search for these processes, and thank
P.~Artoisenet  and J.-P.~Lansberg for many useful discussions and help with theoretical inputs into the {\sc Madgraph}
simulation.

We thank the Fermilab staff and the technical staffs of the participating
institutions for their vital contributions. This work was supported by the
U.S. Department of Energy and National Science Foundation; the Italian Istituto
Nazionale di Fisica Nucleare; the Ministry of Education, Culture, Sports,
Science and Technology of Japan; the Natural Sciences and Engineering Research
Council of Canada; the National Science Council of the Republic of China; the
Swiss National Science Foundation; the A.P. Sloan Foundation; the
Bundesministerium f\"ur Bildung und Forschung, Germany; the Korean World Class
University Program, the National Research Foundation of Korea; the Science and
Technology Facilities Council and the Royal Society, United Kingdom; the Russian
Foundation for Basic Research; the Ministerio de Ciencia e Innovaci\'{o}n, and
Programa Consolider-Ingenio 2010, Spain; the Slovak R\&D Agency; the Academy of
Finland; the Australian Research Council (ARC); and the EU community Marie Curie
Fellowship Contract No. 302103.

\end{acknowledgments}


\end{document}

%% file: author.tex
\affiliation{Institute of Physics, Academia Sinica, Taipei, Taiwan 11529, Republic of China}
\affiliation{Argonne National Laboratory, Argonne, Illinois 60439, USA}
\affiliation{University of Athens, 157 71 Athens, Greece}
\affiliation{Institut de Fisica d'Altes Energies, ICREA, Universitat Autonoma de Barcelona, E-08193, Bellaterra (Barcelona), Spain}
\affiliation{Baylor University, Waco, Texas 76798, USA}
\affiliation{Istituto Nazionale di Fisica Nucleare Bologna, \ensuremath{^{jj}}University of Bologna, I-40127 Bologna, Italy}
\affiliation{University of California, Davis, Davis, California 95616, USA}
\affiliation{University of California, Los Angeles, Los Angeles, California 90024, USA}
\affiliation{Instituto de Fisica de Cantabria, CSIC-University of Cantabria, 39005 Santander, Spain}
\affiliation{Carnegie Mellon University, Pittsburgh, Pennsylvania 15213, USA}
\affiliation{Enrico Fermi Institute, University of Chicago, Chicago, Illinois 60637, USA}
\affiliation{Comenius University, 842 48 Bratislava, Slovakia; Institute of Experimental Physics, 040 01 Kosice, Slovakia}
\affiliation{Joint Institute for Nuclear Research, RU-141980 Dubna, Russia}
\affiliation{Duke University, Durham, North Carolina 27708, USA}
\affiliation{Fermi National Accelerator Laboratory, Batavia, Illinois 60510, USA}
\affiliation{University of Florida, Gainesville, Florida 32611, USA}
\affiliation{Laboratori Nazionali di Frascati, Istituto Nazionale di Fisica Nucleare, I-00044 Frascati, Italy}
\affiliation{University of Geneva, CH-1211 Geneva 4, Switzerland}
\affiliation{Glasgow University, Glasgow G12 8QQ, United Kingdom}
\affiliation{Harvard University, Cambridge, Massachusetts 02138, USA}
\affiliation{Division of High Energy Physics, Department of Physics, University of Helsinki, FIN-00014, Helsinki, Finland; Helsinki Institute of Physics, FIN-00014, Helsinki, Finland}
\affiliation{University of Illinois, Urbana, Illinois 61801, USA}
\affiliation{The Johns Hopkins University, Baltimore, Maryland 21218, USA}
\affiliation{Institut f\"{u}r Experimentelle Kernphysik, Karlsruhe Institute of Technology, D-76131 Karlsruhe, Germany}
\affiliation{Center for High Energy Physics: Kyungpook National University, Daegu 702-701, Korea; Seoul National University, Seoul 151-742, Korea; Sungkyunkwan University, Suwon 440-746, Korea; Korea Institute of Science and Technology Information, Daejeon 305-806, Korea; Chonnam National University, Gwangju 500-757, Korea; Chonbuk National University, Jeonju 561-756, Korea; Ewha Womans University, Seoul, 120-750, Korea}
\affiliation{Ernest Orlando Lawrence Berkeley National Laboratory, Berkeley, California 94720, USA}
\affiliation{University of Liverpool, Liverpool L69 7ZE, United Kingdom}
\affiliation{University College London, London WC1E 6BT, United Kingdom}
\affiliation{Centro de Investigaciones Energeticas Medioambientales y Tecnologicas, E-28040 Madrid, Spain}
\affiliation{Massachusetts Institute of Technology, Cambridge, Massachusetts 02139, USA}
\affiliation{University of Michigan, Ann Arbor, Michigan 48109, USA}
\affiliation{Michigan State University, East Lansing, Michigan 48824, USA}
\affiliation{Institution for Theoretical and Experimental Physics, ITEP, Moscow 117259, Russia}
\affiliation{University of New Mexico, Albuquerque, New Mexico 87131, USA}
\affiliation{The Ohio State University, Columbus, Ohio 43210, USA}
\affiliation{Okayama University, Okayama 700-8530, Japan}
\affiliation{Osaka City University, Osaka 558-8585, Japan}
\affiliation{University of Oxford, Oxford OX1 3RH, United Kingdom}
\affiliation{Istituto Nazionale di Fisica Nucleare, Sezione di Padova, \ensuremath{^{kk}}University of Padova, I-35131 Padova, Italy}
\affiliation{University of Pennsylvania, Philadelphia, Pennsylvania 19104, USA}
\affiliation{Istituto Nazionale di Fisica Nucleare Pisa, \ensuremath{^{ll}}University of Pisa, \ensuremath{^{mm}}University of Siena, \ensuremath{^{nn}}Scuola Normale Superiore, I-56127 Pisa, Italy, \ensuremath{^{oo}}INFN Pavia, I-27100 Pavia, Italy, \ensuremath{^{pp}}University of Pavia, I-27100 Pavia, Italy}
\affiliation{University of Pittsburgh, Pittsburgh, Pennsylvania 15260, USA}
\affiliation{Purdue University, West Lafayette, Indiana 47907, USA}
\affiliation{University of Rochester, Rochester, New York 14627, USA}
\affiliation{The Rockefeller University, New York, New York 10065, USA}
\affiliation{Istituto Nazionale di Fisica Nucleare, Sezione di Roma 1, \ensuremath{^{qq}}Sapienza Universit\`{a} di Roma, I-00185 Roma, Italy}
\affiliation{Mitchell Institute for Fundamental Physics and Astronomy, Texas A\&M University, College Station, Texas 77843, USA}
\affiliation{Istituto Nazionale di Fisica Nucleare Trieste, \ensuremath{^{rr}}Gruppo Collegato di Udine, \ensuremath{^{ss}}University of Udine, I-33100 Udine, Italy, \ensuremath{^{tt}}University of Trieste, I-34127 Trieste, Italy}
\affiliation{University of Tsukuba, Tsukuba, Ibaraki 305, Japan}
\affiliation{Tufts University, Medford, Massachusetts 02155, USA}
\affiliation{University of Virginia, Charlottesville, Virginia 22906, USA}
\affiliation{Waseda University, Tokyo 169, Japan}
\affiliation{Wayne State University, Detroit, Michigan 48201, USA}
\affiliation{University of Wisconsin, Madison, Wisconsin 53706, USA}
\affiliation{Yale University, New Haven, Connecticut 06520, USA}

\author{T.~Aaltonen}
\affiliation{Division of High Energy Physics, Department of Physics, University of Helsinki, FIN-00014, Helsinki, Finland; Helsinki Institute of Physics, FIN-00014, Helsinki, Finland}
\author{S.~Amerio\ensuremath{^{kk}}}
\affiliation{Istituto Nazionale di Fisica Nucleare, Sezione di Padova, \ensuremath{^{kk}}University of Padova, I-35131 Padova, Italy}
\author{D.~Amidei}
\affiliation{University of Michigan, Ann Arbor, Michigan 48109, USA}
\author{A.~Anastassov\ensuremath{^{w}}}
\affiliation{Fermi National Accelerator Laboratory, Batavia, Illinois 60510, USA}
\author{A.~Annovi}
\affiliation{Laboratori Nazionali di Frascati, Istituto Nazionale di Fisica Nucleare, I-00044 Frascati, Italy}
\author{J.~Antos}
\affiliation{Comenius University, 842 48 Bratislava, Slovakia; Institute of Experimental Physics, 040 01 Kosice, Slovakia}
\author{G.~Apollinari}
\affiliation{Fermi National Accelerator Laboratory, Batavia, Illinois 60510, USA}
\author{J.A.~Appel}
\affiliation{Fermi National Accelerator Laboratory, Batavia, Illinois 60510, USA}
\author{T.~Arisawa}
\affiliation{Waseda University, Tokyo 169, Japan}
\author{A.~Artikov}
\affiliation{Joint Institute for Nuclear Research, RU-141980 Dubna, Russia}
% ...

\collaboration{CDF Collaboration}
\altaffiliation[With visitors from]{
\ensuremath{^{a}}University of British Columbia, Vancouver, BC V6T 1Z1, Canada,
\ensuremath{^{b}}Istituto Nazionale di Fisica Nucleare, Sezione di Cagliari, 09042 Monserrato (Cagliari), Italy,
\ensuremath{^{c}}University of California Irvine, Irvine, CA 92697, USA,
\ensuremath{^{d}}Institute of Physics, Academy of Sciences of the Czech Republic, 182~21, Czech Republic,
\ensuremath{^{e}}CERN, CH-1211 Geneva, Switzerland,
\ensuremath{^{f}}Cornell University, Ithaca, NY 14853, USA,
\ensuremath{^{g}}University of Cyprus, Nicosia CY-1678, Cyprus,
\ensuremath{^{h}}Office of Science, U.S. Department of Energy, Washington, DC 20585, USA,
\ensuremath{^{i}}University College Dublin, Dublin 4, Ireland,
\ensuremath{^{j}}ETH, 8092 Z\"{u}rich, Switzerland,
\ensuremath{^{k}}University of Fukui, Fukui City, Fukui Prefecture, Japan 910-0017,
\ensuremath{^{l}}Universidad Iberoamericana, Lomas de Santa Fe, M\'{e}xico, C.P. 01219, Distrito Federal,
\ensuremath{^{m}}University of Iowa, Iowa City, IA 52242, USA,
\ensuremath{^{n}}Kinki University, Higashi-Osaka City, Japan 577-8502,
\ensuremath{^{o}}Kansas State University, Manhattan, KS 66506, USA,
\ensuremath{^{p}}Brookhaven National Laboratory, Upton, NY 11973, USA,
\ensuremath{^{q}}Istituto Nazionale di Fisica Nucleare, Sezione di Lecce, Via Arnesano, I-73100 Lecce, Italy,
\ensuremath{^{r}}Queen Mary, University of London, London, E1 4NS, United Kingdom,
\ensuremath{^{s}}University of Sydney, NSW 2006, Australia,
\ensuremath{^{t}}Muons, Inc., Batavia, IL 60510, USA,
\ensuremath{^{u}}Nagasaki Institute of Applied Science, Nagasaki 851-0193, Japan,
\ensuremath{^{v}}National Research Nuclear University, Moscow 115409, Russia,
\ensuremath{^{w}}Northwestern University, Evanston, IL 60208, USA,
\ensuremath{^{x}}University of Notre Dame, Notre Dame, IN 46556, USA,
\ensuremath{^{y}}Universidad de Oviedo, E-33007 Oviedo, Spain,
\ensuremath{^{z}}CNRS-IN2P3, Paris, F-75205 France,
\ensuremath{^{aa}}Universidad Tecnica Federico Santa Maria, 110v Valparaiso, Chile,
\ensuremath{^{bb}}The University of Jordan, Amman 11942, Jordan,
\ensuremath{^{cc}}Universite catholique de Louvain, 1348 Louvain-La-Neuve, Belgium,
\ensuremath{^{dd}}University of Z\"{u}rich, 8006 Z\"{u}rich, Switzerland,
\ensuremath{^{ee}}Massachusetts General Hospital, Boston, MA 02114 USA,
\ensuremath{^{ff}}Harvard Medical School, Boston, MA 02114 USA,
\ensuremath{^{gg}}Hampton University, Hampton, VA 23668, USA,
\ensuremath{^{hh}}Los Alamos National Laboratory, Los Alamos, NM 87544, USA,
\ensuremath{^{ii}}Universit\`{a} degli Studi di Napoli Federico I, I-80138 Napoli, Italy
}
\noaffiliation